\documentclass[conference]{IEEEconf}

\usepackage{paralist}
\usepackage{listings}
\usepackage{xspace}
\usepackage{hyperref}
\usepackage{wrapfig}

\usepackage{tikz}
\usetikzlibrary{shapes,arrows}
\usetikzlibrary{calc}
\usetikzlibrary{fit,positioning}
\usetikzlibrary{patterns}

\input epsf
\usepackage{graphicx}

\newcommand{\Peass}{Peass\xspace}
\newcommand{\Kieker}{Kieker\xspace}
\newcommand{\PercentJMH}{3.2\,\%\xspace}
\newcommand{\PercentPeass}{58.7\,\%\xspace}
\newcommand{\PercentUnitTest}{26.7\,\%\xspace}
\newcommand{\PercentNotCovered}{14.6\,\%\xspace}

\begin{document}

\title{\textbf{\Large Automated Identification of Performance Changes at Code Level\\}}

\author{David Georg Reichelt$^{1,2,*}$, Stefan Kühne$^{1}$, and Wilhelm Hasselbring$^{3}$\\
	\normalsize $^{1}$Universität Leipzig, Leipzig, Saxony, Germany, $^{2}$Lancaster University Leipzig, Leipzig, Saxony, Germany,\\
	\normalsize $^{3}$Christian-Albrechts-Universität zu Kiel, Kiel, Schleswig-Holstein, Germany\\
	\normalsize d.g.reichelt@lancaster.ac.uk, stefan.kuehne@uni-leipzig.de, hasselbring@email.uni-kiel.de\\
	\normalsize *corresponding author
}


\maketitle
\begin{abstract}
To develop software with optimal performance, even small performance changes need to be identified. Identifying performance changes is challenging since the performance of software is influenced by non-deterministic factors. Therefore, not every performance change is measurable with reasonable effort. In this work, we discuss which performance changes are measurable at code level with reasonable measurement effort and how to identify them. We present (1) an analysis of the boundaries of measuring performance changes, (2) an approach for determining a configuration for reproducible performance change identification, and (3) an evaluation comparing of how well our approach is able to identify performance changes in the application server Jetty compared with the usage of Jetty's own performance regression benchmarks.\\
Thereby, we find (1) that small performance differences are only measurable by fine-grained measurement workloads, (2) that performance changes caused by the change of one operation can be identified using a unit-test-sized workload definition and a suitable configuration, and (3) that using our approach identifies small performance regressions more efficiently than using Jetty's performance regression benchmarks.
\end{abstract}
\IEEEoverridecommandlockouts

%
\IEEEpeerreviewmaketitle

\section{Introduction}
\label{sec:introduction}

During development and maintenance of software, we often intend to assure that the application performance is optimal. Regardless of whether we strive for optimality regarding response time or resource usage, optimal performance behavior requires efficient implementations at source code level. More efficient implementations at code level differ from less efficient implementations in terms of the selection of algorithms, definition of data structures and usage of library APIs. 
To assure that the software created by a changing code base contains efficient implementations at code level, we need to detect performance changes caused by code changes. Since active software projects create several commits a day and developers do not remember the details of old commits, performance changes should be detected by measurement at least daily. This makes it possible to understand whether optimizations have the positive effect the developers expected and to fix regressions whenever possible. 
The detection of performance changes using code level measurement\footnote{The term \textit{code level measurement} in this paper refers to the measurement by execution of both, workload inducing and measurement code, inside the program itself (as mostly done by benchmarks). Code level measurement mostly aims at comparing implementations of code inside one module. The opposite is architecture level measurement, which is done by load tests and monitoring. Architecture-level measurement mostly aims for comparing the effects of architectural or deployment decisions.}
is challenging since performance measurements are influenced by non-deterministic factors like Just-In-Time-Compilation of virtual machines\footnote{In this paper, virtual machines refer to the runtime environment making bytecode portable on different plattforms, e.g. the JVM or the .NET runtime. It does not relate to the term virtual machines as used in virtualization of operating systems.} (VMs), memory fragmentation on program start and CPU scaling \cite{georges2007statistically}. While these effects are similiar in different types of virtual machines,\footnote{For example in Javascript \cite{selakovic2016performance}, Pharo \cite{alcocer2015} and the .NET-Runtime \cite{quiroga2016optimizing}, similiar measurement processes are used. Compiled or solely interpreted languages use different measurement processes.} we describe how to measure performance changes in the Java Virtual Machine (JVM).


Due to these non-deterministic effects, performance changes can only be identified with some uncertainty. Since we only have limited measurement resources available, especially if we want to identify the performance changes of a day, it is only possible to identify performance changes of a certain size. Therefore, we discuss how the relation between the size of a performance change in terms of measured run time and the standard deviation of measurements affects the required number of measurements. We find that performance changes are only measurable with reasonable effort if the relative size of a change is at least half the standard deviation. Therefore, if our test case has a run time of one second and the standard deviation is 0.1\,s (10\,\%), a change of 0.01\,s (1\,\%) is not measurable with reasonable effort. 

This implies that detecting small performance changes--i.e.\ performance changes that are measurable at code level--using load tests of the whole application is not possible. While load tests and coarse-grained monitoring are sufficient for the detection of performance problems that are rooted at the architecture or deployment level, the detection of performance changes at the code level requires finer-grained measurement workloads. Since performance changes at the code level are hardly measurable by load tests, a performance change at code level might not influence the perceived performance significantly. Nevertheless, many of those changes might pile up and change the performance behavior of the software. Therefore, understanding them is a huge benefit when striving for optimal application performance.

Creating and maintaining smaller benchmarks requires significant development effort. One option to create smaller benchmarks without significant development effort is the automated transformation of existing unit tests. This approach is pursued by several authors \cite{reichelt2019asedemo,eidDetecting2020,chen2019analyzing}. In our previous work \cite{reichelt2019asedemo}, we developed the \Peass approach, which measures performance changes of unit tests and contains a regression test selection for measurement speedup and a root cause analysis. In this work, we ground the performance measurement of \Peass, which is the most crucial part, through a statistical analysis of measurability of performance changes.

To be able to identify performance regressions in unit-test-sized workloads, we answer the research question \textbf{(RQ~I)}: How can performance changes be identified at code level? We find that, for unit-test-sized workloads running in the JVM, small performance changes can only be identified with a certain minimum parameters, e.g. if at least 30 VM starts are executed. The parallelized execution of one VM start for both versions does not hinder the identification of performance changes. The parameters and the decision whether to start VMs in parallel might need adaption for workloads that differ in size or resource demand.

To evaluate the effectiveness of the measurement of performance changes at code level by transformed unit tests, we address research question \textbf{(RQ~II)}: Which share of performance changes can be identified by measuring the unit test performance and by existing performance regression benchmarks? This is accomplished via injection of artificial performance regressions into the Jetty application server. We find that \PercentPeass of the changes are identified by measuring the transformed unit tests and that \PercentJMH of the changes are identified by Jetty's performance regression benchmarks, i.e.\ that unit test measurement is able to identify more performance regressions than existing benchmarks.

The remainder of this paper is organized as follows: First, we describe the boundaries of performance change detection in Section~\ref{kap:boundaries}. In Section~\ref{kap:detection}, we discuss the configuration of performance change detection. The described method for automated performance change identification is evaluated in Section~\ref{kap:evaluation}. In Section~\ref{kap:related}, we discuss related work. Finally, in Section~\ref{kap:summary} we give a summary and outlook.

\section{Boundaries of Detecting Performance Changes}
\label{kap:boundaries}

In this section, we describe the general process for performance measurement in managed runtimes like Java. Afterwards, we determine the typical relative standard deviation of performance measurement workloads and infer statistical boundaries of performance change detection.

\subsection{General Process}

The measurement of the performance of a software is influenced by non-deterministic factors. At the hardware and operating system level, these factors include
\begin{inparaenum}[(1)]
  \item memory fragmentation on program start,
  \item CPU scaling and temperature during execution and
  \item parallel processes in operating systems, e.g. checks for updates.
\end{inparaenum}

Inside VMs, the non-deterministic factors influencing performance additionally include
\begin{inparaenum}[(1)]
  \item garbage collections, occurring in a non-foreseeable manner and slowing down the program,
  \item thread scheduling, eventually choosing different execution orders in different VM starts and
  \item just-in-time compilations and optimization.
\end{inparaenum} 
Just-in-time compilation and optimizations can influence long-term performance due to the different states the compilation and optimization may end up in \cite{georges2007statistically}.

To measure performance with statistical rigor, Georges et al.~\cite{georges2007statistically} recommend to run several VM starts (VMs\footnote{In parts of the literature and tools, e.g. \textsc{jmh}, these VM starts are called forks. We use the terminology of Georges et al.~\cite{georges2007statistically} and call the count of the VM starts for measurement VMs.}). These VM executions contain reruns of the workload. These reruns are split into two parts: The initial runs to warm up a VM (warmup\footnote{To avoid late optimization of the measurement code itself, measurements are also executed during the warmup and discared afterwards.} iterations) and the reruns inside this VM for the measurement (iterations). Afterwards, they propose to compare measurement results using comparison of the confidence intervals of the mean values of the measurement iterations.

For practical execution of performance benchmarks there exists a variety of tools, including \textsc{jmh}\footnote{Source of \textsc{jmh}: \url{https://github.com/openjdk/jmh}}, JUnitPerf\footnote{Source of JUnitPerf: \url{https://github.com/clarkware/junitperf}} and JUnitBench.\footnote{Source of JUnitBench: \url{https://github.com/tpounds/junitbench}} These tools provide default parameters and give the user the option to specify the VM count and the length of warmup and measurement iterations. This length might be specified in terms of workload executions or measurement duration depending on the tool.

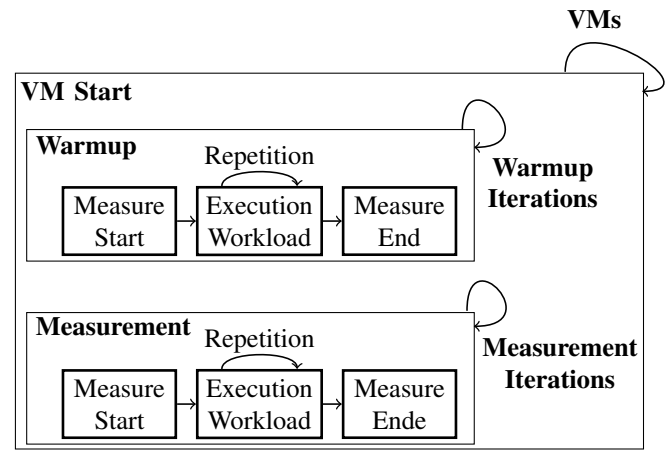
\begin{figure}[h]
  \centering
  \begin{tikzpicture}[node distance = 0.25cm, auto, every node/.style={line width=1pt,draw,shape=rectangle,minimum width = 1.5cm,align=center}]
  \node (Workload1) {Execution\\ Workload};
	\node[left=of Workload1] (Startzeit1) {Measure\\ Start};  
	\node[right=of Workload1] (Endzeit1) {Measure\\ End};
	\path[draw, ->, semithick] (Startzeit1) -- (Workload1);
	\path[draw, ->, semithick] (Workload1) -- (Endzeit1);
	    
	\path[->, semithick]([xshift=-0.5cm]Workload1.north) edge[out = 90, in = 90,distance=0.25cm] node[draw=none, above=-0.1cm] {Repetition} ([xshift=0.5cm]Workload1.north);

	  \draw[rounded corners=0cm] ($(Workload1.north west) - (2.25, -0.75cm)$) rectangle ($(Workload1.north west) - (-3.7cm, 1.0cm)$) node [draw=none,pos=.3,xshift=-1.8cm] (warmup){\textbf{Warmup}};    
	  
	  \path[->, semithick]([xshift=5.00cm,yshift=-0.05cm]warmup.north) 
	    edge[out = 90, in = 0,distance=1cm] 
	    node[draw=none, xshift=-0.5cm,below right=0.5cm and 0.25cm] {\textbf{Warmup}\\\textbf{Iterations}} 
	    ([xshift=4.35cm]warmup.east);

	  \node[below=1.5cm of Workload1] (Workload2) {Execution\\ Workload};
	  \node[left=of Workload2] (Startzeit2) {Measure\\ Start};  
	  \node[right=of Workload2] (Endzeit2) {Measure\\ Ende};
	  \path[draw, ->, semithick] (Startzeit2) -- (Workload2);
	  \path[draw, ->, semithick] (Workload2) -- (Endzeit2);

	  \path[->, semithick]([xshift=-0.5cm]Workload2.north) edge[out = 90, in = 90,distance=0.25cm] node[draw=none, above=-0.1cm] {Repetition} ([xshift=0.5cm]Workload2.north);
	  
	  \draw[rounded corners=0cm] ($(Workload2.north west) - (2.25, -0.75cm)$) rectangle ($(Workload2.north west) - (-3.7cm, 1.0cm)$) node [draw=none,pos=.25,xshift=-1.5cm] (messung){\textbf{Measurement}};   
	
	  \path[->, semithick]([xshift=4.70cm,yshift=-0.05cm]messung.north) 
	    edge[out = 90, in = 0,distance=1cm] 
	    node[draw=none, xshift=-0.6cm,below right=0.5cm and 0.25cm] {\textbf{Measurement}\\\textbf{Iterations}} 
	    ([xshift=3.625cm]messung.east);     

	  \draw[rounded corners=0cm] ($(Workload1.north west) - (2.4, -1.5cm)$) rectangle ($(Workload1.north west) - (-5.95cm, 3.5cm)$) node [draw=none,pos=.1,xshift=-0.9cm] (vmstart){\textbf{VM Start}};  
	  
	\path[->, semithick]([xshift=6.5cm,yshift=0.0cm]vmstart.north) 
	    edge[out = 90, in = 0,distance=1cm] 
	    node[draw=none, xshift=-0.5cm,above=0.20cm] {\textbf{VMs}} 
	    ([xshift=6.65cm]vmstart.east);   
  \end{tikzpicture}
  \caption{General Measurement Process}
  \label{fig:measurementProcess}
\end{figure}

To reduce the monitoring overhead, some monitoring tools repeat the workload inside of one iteration, e.g. \textsc{jmh} repeats the workload for a given time. Therefore, we split the workload execution inside of one VM in two steps: For $iteration$ times, we measure the start time, execute the workload $repetitions$ times and measure the iteration end time. Subsequently, we calculate the duration per repetition by dividing the average difference of start and end time by the count of repetitions. This reduces measurement overhead and increases accuracy of the measured values. Figure~\ref{fig:measurementProcess} summarizes the described measurement process.

\subsection{Standard Deviation of Workloads}

To detect a performance change in a software, we need to measure the performance of a defined workload in the two examined versions and infer whether there has been a performance change. This can be done using load tests, which measure the performance of the whole application or the performance of exposed technical interfaces, e.g., by sending HTTP requests. Alternatively, performance measurement can be done using benchmarks, which may exercise source code of whole components or of smaller parts of the software. Due to the aforementioned non-determinism, performance measurements are subject to deviations. These deviations vary with the size of the workload in terms of execution duration, which implies that it is harder to detect a small performance change by a load test, which measures the overall software, than by a benchmark only measuring a small part of the software. In this subsection, we determine the relative standard deviation of such benchmarks or load tests. Since the duration cannot be derived statically, we determine the standard deviation in relation to the call count of equally sized method executions.

To determine the standard deviation of real workloads, representative workloads for software are required that use different parts of the system. 
Therefore, we use the following three workload types with the size $s$:
\begin{inparaenum}[(1)]
  \item Addition of $s$ random numbers, (\lstinline'Add'-workload)
  \item Reservation of $s$ arrays consisting of 3 \lstinline'int's and
  \item Generating $s$ random numbers and printing them to \lstinline'System.out'.
\end{inparaenum}
These workload types have already been used in our previous work for performance measurement calibration \cite{reichelt2018wospc}. They are used because they utilize parts of the hardware which may have different characteristics regarding performance, e.\,g.\,RAM-intense workloads may have different standard deviation or warmup duration than CPU-intense workload. Since hard disk interaction is influenced by other factors, such as write buffers and specifics of the particular hard disk, we omit hard disk usage workloads. 

\begin{figure}
  \includegraphics[width=\columnwidth]{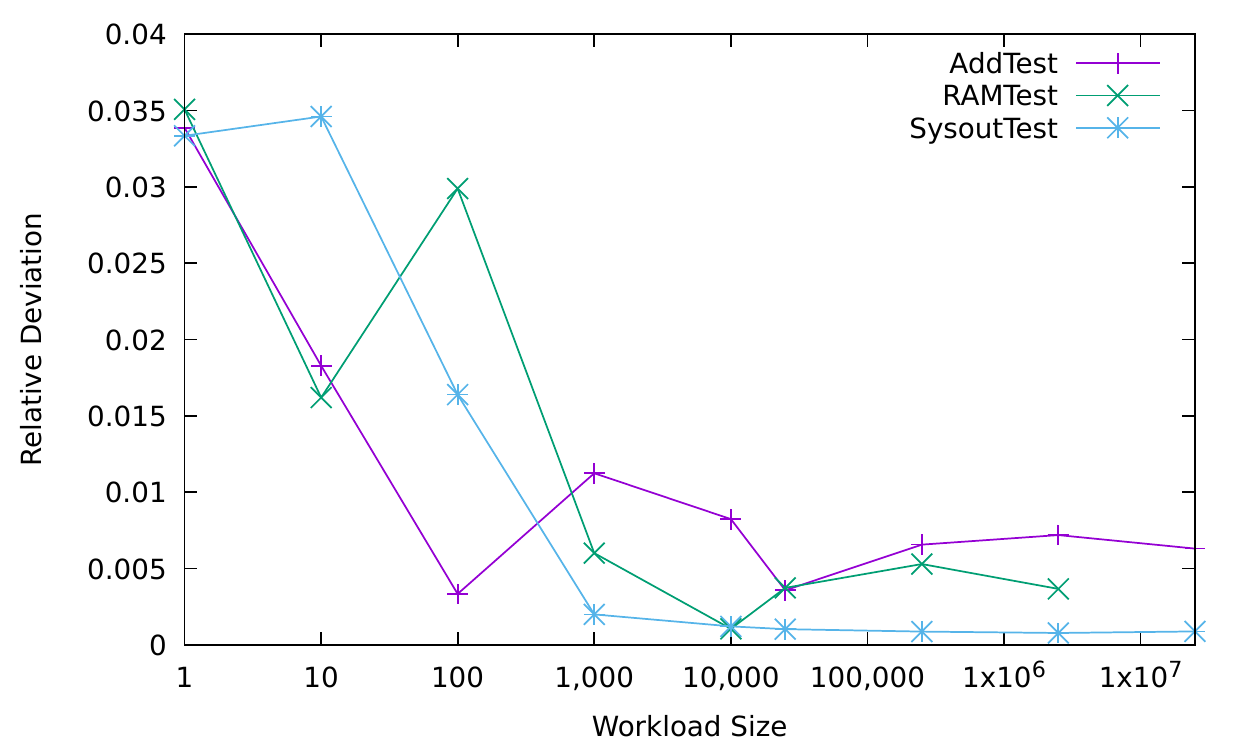}
  \caption{Evolution of the Relative and Absolute Standard Deviation with Workload Size Increase}
  \label{fig:graphMeasurements}
\end{figure}

We executed all presented measurements in this paper, if not otherwise specified, on i7-4770 CPU with 3.40\,GHz, 16\,GB RAM, Ubuntu 20.04 and OpenJDK 1.8.0\_275 (defaulting to 25\,\% heap of RAM size, i.e. 4\,GB). For the chosen workloads, the standard deviation of different workload sizes is depicted in Figure \ref{fig:graphMeasurements}.\footnote{Data are available in \lstinline'size-evolution.tar' in \url{https://zenodo.org/record/6427379\#.YnKUbVxBxhE}} The RAMTest could not be executed on more than 1,000,000 executions because of an \lstinline'OutOfMemoryError'. This shows that the relative standard deviation $\sigma/\mu$ stays between \textbf{0.1\,\% and 1\,\%} when the workload size is varying, except for very small workload sizes. For these small worload sizes, the relative standard deviation is below 4\,\%.

\subsection{Statistical Boundaries of Detecting Changes}

The boundaries of detection of performance changes are determined by the population effect size $\gamma$ and the count of VM measurements: A bigger performance change is easier to measure and if we repeat the measurement more often, changes are identified with higher precision. 

Given measurements from two distributions representing two versions of software $X$ and $Y$ and the pooled standard deviation $\sigma_S$, the population effect size is $\gamma=\frac{\mu_X-\mu_Y}{\sigma_S}$, i.e. the effect size our experiment would converge to. Since the relative standard deviation of workloads does not decrease with increasing workload size, the population effect size is smaller for big workloads if the same absolute changes appear. 

\begin{figure}
  \includegraphics[width=\columnwidth]{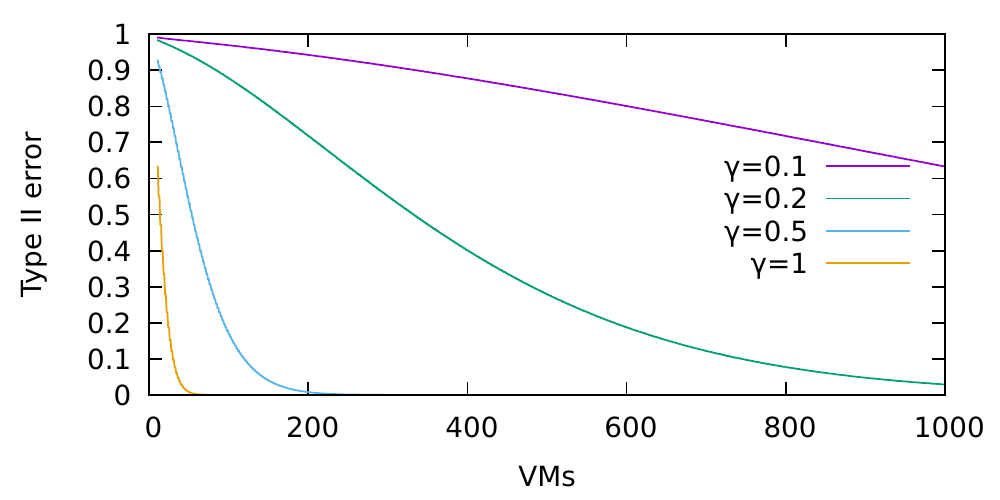}
  \caption{Evolution of the Type~II error with Count of VM Starts (Significance 99\,\%)}
  \label{fig:graphStatistic}
\end{figure}

The relation of effect size $\gamma$, sample size (in our case: count of VM starts $VMs$), Type~I error $\alpha$ (the limiting value the false positive rate converges against) and Type~II error $\beta$ (the limiting value the false negative rate converges against) is well-known \cite{cohen1970approximate}. The Type II error for two-sided tests is determined by $_uz_{1-\beta}=\gamma*\sqrt{VMs/2}-z_{1-\alpha/2}$. We assume that the Type~I error should be below 1\,\%, i.e. we do not allow more than one false positive measurements in 100 measurements. Based on the population effect size, we can therefore derive the Type~II errors depicted in Figure~\ref{fig:graphStatistic}. While analyzing performance measurements with a difference bigger than standard deviation ($\gamma \geq 1$) results in a low Type~II error with a low count of VMs, measuring differences smaller than the standard deviation ($\gamma<1$) requires many VMs. 

\paragraph{Example} 
If in a workload consisting of 1,000 equal sized method calls ten calls are added, the performance changes by $\approx$0.5\,\%. 30 VMs are considered as a large sample size \cite{georges2007statistically}. If we use 30\,VMs, the change of 0.5\,\% (with 0.5\,\% relative standard deviation, $\gamma=1$) is measurable with Type~II error of 9.73\,\% (yellow curve, 30\,VMs). Missing every 10th performance change is not acceptable. If we use 50\,VMs instead, the Type~II error is 0.76\,\%, which is acceptable. If only one operation is added, the performance changes by $\approx$0.1\,\%, the effect size is $\gamma=0.2$, which would result in a Type~II error of 94.2\,\% for 50\,VMs, i.e. the change not detected in 19 out of 20 measurements.

\paragraph{Implications} 
We intend to keep our Type~II below 1\,\%, since we also do not accept more than one false negative in 100 measurements. With this prerequisites, we need to execute a sufficient count of VMs, which depends on our population effect size $\gamma$. 
VM executions for a workload size of 1,000 take at least 97 seconds in our measurements, if we use 10,000,000 repetitions of the workload. Usually, when comparing the performance of fine-grained tests, measurement is not allowed to take more than half a day; otherwise, the developers would not get fast enough feedback to fix performance regressions for their commits on the next day. Therefore, executing more than $\approx$~891\,VMs ($=12*60*60/97$) would not be feasible for those workloads. For bigger workloads, even fewer VM executions are possible.

Due to the limited number of available VM executions for regular measurements and our requirements regarding Type~I and Type~II error, only performance changes with a certain size are measurable. Depending on the execution time, this amount of possible VM executions might vary. Nevertheless, an effect size of $\gamma=0.1$, which would require 4,808\,VM executions, cannot be measured with feasible effort, while an effect size of $\gamma=0.5$ will be measurable under most circumstances.
Therefore, for detecting small performance changes, fine-grained measurement data are needed. A top-level measurement of the performance of a load test or a benchmark covering a big workload is not sufficient. Also, the use of monitoring frameworks like OpenTelemetry\footnote{https://opentelemetry.io/} or Kieker \cite{Kieker2020} will only be able to detect small performance changes if the duration of small workloads is measured. Therefore, for identification of performance changes outside of production environments, small workload definitions are required. 


\section{Detection of Performance Changes} 
\label{kap:detection}

In this section, we answer \textbf{RQ~I} by describing our method for the detection of performance changes. In the first subsection, we describe our approach for configuration of the performance measurement. For our approach, we need typical workload sizes; the determination of these workload sizes is described in the second subsection. In the third subsection, we describe our obtained measurement configuration.

\subsection{Approach}

For the examination of performance changes at code level, fine-grained measurements are needed. Following the \textit{unit test assumption},  ``the performance of relevant use cases of a program correlates with the performance of at least a part of its unit tests, if the performance is not driven mainly by external factors'' \cite{reichelt2018wospc}. Therefore, the workloads of some unit tests can be used for identification of performance changes at code level. 

This does not apply to all unit tests, e.g. some tests might measure corner cases and therefore not be representative of typical interactions with the defined interface. Therefore, some filtering of the tests needs to be done. This could happen based on monitoring data of production systems that contain execution traces. Unit tests, that only test corner cases, would call methods that are rare in the execution traces. This work focuses on the measurement process, hence the filtering is out of scope of this work. Accordingly, the measurement configuration we obtain can be used for unit test sized workloads not using external calls, like calls to temporary databases, having the same order of magnitude of their workloads size.

To compare different configurations, we first describe the possible configurations and afterwards the method for determination of the $F_1$-score of each configuration. Finally, we describe how we select the best configuration based on the $F_1$-scores.


\paragraph{Configurations}
To measure performance changes by workloads of unit tests, the unit tests need to be transformed to contain measurement iterations inside of each started VM. Afterwards, different configurations need to be set, including measurement parametrization, the technical measurement environment and the configuration of the analysis.

\textbf{Measurement Parametrization}: Which iteration and VM count should be used, how often the workload should be repeated between measurement start and stop and whether garbage collection should be triggered between two iterations.

\textbf{Technical Measurement Environment}: In which technical environment the measurement process should be started, e.g. whether measurement can be parallelized on the same machine or whether standard output may be ignored.

\textbf{Analysis Configuration}: Which analysis configuration should be used. This includes which statistical test should be used, e.g. t-test, confidence interval comparison or Mann-Whitney-test, how it should be parameterized (e.g. significance level) and what preprocessing steps should be done (i.e. whether to execute outlier removal).

\paragraph{$F_1$-Score Determination} 
The accuracy of change identification is measured by the $F_1$-score, i.e. the harmonic mean of precision (true positive / (true positive + false positive)) and recall (true positive / (true positive + false negative); 100\,\% means that all performance changes are detected correctly and 0\,\% means that no performance change was detected correctly. 

For determination of the $F_1$-score of a performance measurement and analysis method, we need workloads where we can control the effect size of the change. Since the already described workloads of \cite{reichelt2018wospc} allow the specification of the workload size, we reuse them. We execute each artificial workload with the sizes $s$ and $s+d$, the configuration we intend to examine, and the maximum of measurement iterations and VM starts we want to examine. Afterwards, we check for given iteration counts $i$ and VM start counts $v$ to how well they identify the performance change. 

For each configuration with $v$ VMs and $i$ iterations, we randomly sample $v$ VMs and select the first $i$ iterations
\begin{inparaenum}[(1)]
  \item from both measurements of both versions and check whether the performance change is identified correctly, and, 
  \item from measurements of first version and check whether the equal performance is correctly identified. 
\end{inparaenum}
The sampling and analysis is repeated 10,000 times, so our $F_1$-score can be calculated from true and false positives. Based on the $F_1$-score and measurement duration, we select the appropriate measurement method.

\paragraph{Selection of Best Configuration}
When comparing different measurement configurations and statistical tests, we want an $F_1$-score of at least 99\,\%. While increasing the VM count always increases the $F_1$-score, the same does not hold for the iteration count. More iterations may result in more or different optimizations or compilations done by the JVM during the execution. The optimizations, the compilations, or the optimization and compilation processes themself might lead to a (temporarily) higher standard deviation, which decreases the $F_1$-score. To obtain the warmed-up performance, the measurement needs to be repeated until no decreased $F_1$-score is appearing due to higher iteration count.

Furthermore, the execution duration for the measurement is also crucial. Given the same product $iteration*repetition$, i.e. same overall execution count, higher repetition counts result in lower execution times. Therefore, we search for the lowest VM and $iteration*repetition$ count combination with the following properties: 
\begin{inparaenum}
  \item We want at maximum one false positive or one false negative for 100 measurements, so the $F_1$-score should be at least 99\,\%.
  \item An increase in iteration count should not lead to a reduction of the $F_1$ score, so we do not mistakenly classify optimization or compilation influences for performance changes.
  \item If two parameter combinations meet both requirements, the combination with higher repetition count is used.
\end{inparaenum}



\subsection{Workload Size}

The artifical workloads require a specification of the workload size, i.e. the count of operations that should be executed. To determine the relevant workload size, we use the typical size of unit tests in terms of lines of code (excluding non-statement code, e.g. comments) as a proxy. 

\begin{figure}[h]
  \includegraphics[width=\columnwidth]{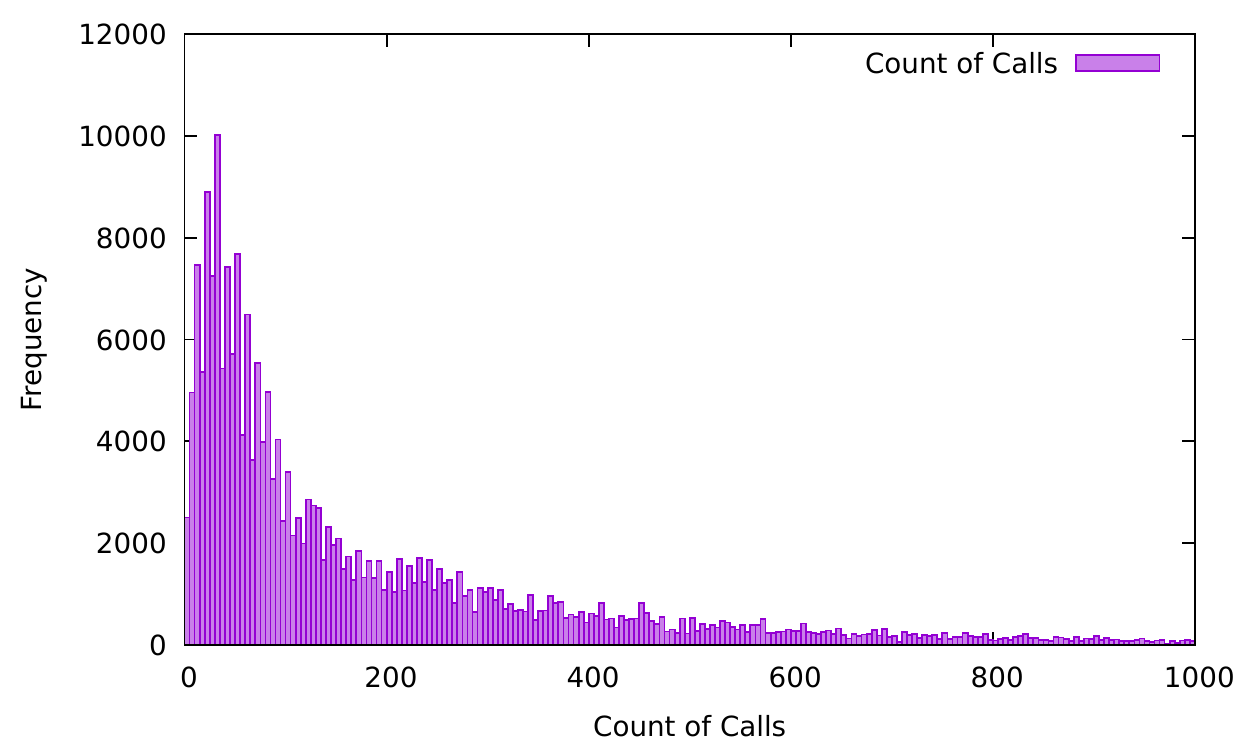}
  \caption{Operation Call Counts of Unit Tests}
  \label{fig:callCount}
\end{figure}

To identify the typical size of unit tests, we gather the workload sizes of Apache Commons projects. Those projects are created to supply reusable Java components for different purposes. Due to their widespread use and big user community, they are a representative example of high quality source code, and therefore we use their typical unit test size. 

Figure~\ref{fig:callCount} shows the count of method calls of unit tests.\footnote{Dataset: \url{https://doi.org/10.5281/zenodo.6517822}} The median of the count of method calls is 126 and the average method contains 2.29 lines excluding comments, getter calls, setter calls, empty lines and non-statement lines like \lstinline'return' and \lstinline'break'. The average method line count of 2.29 is consistent with the recommendation of making methods as small as possible for clean code \cite{martin2008clean}. We therefore make the simplifying assumption that a workload size of 300 method calls is a suitable proxy for measuring real world performance changes at code level. 

Performance changes at code level might cause different performance change sizes in production, since most methods are called more often in production than in the unit test. A minor performance change at code level might therefore cause a major performance change in production. Therefore, we aim for identification of as small performance changes as possible with feasible effort. For the typical unit test size of 300 method calls, the smallest possible performance change is a change of one statement, i.e. a change of $0.\bar{3}\%$. This should be detected by our measurement configuration.

\subsection{Measurement Configuration}

In the following, we exemplarily describe two parts of the configuration: How to choose VM and execution count (part of the \textit{measurement parametrization)}, whether to parallelize the measurements (part of the \textit{technical measurement environment}) and whether to remove outliers (part of the \textit{analysis configuration}). Finally, we discuss the generalizability of our configuration. Our dataset is available.\footnote{Dataset: \url{https://doi.org/10.5281/zenodo.6427379}}

\begin{figure}[h!]
  \includegraphics[width=12cm]{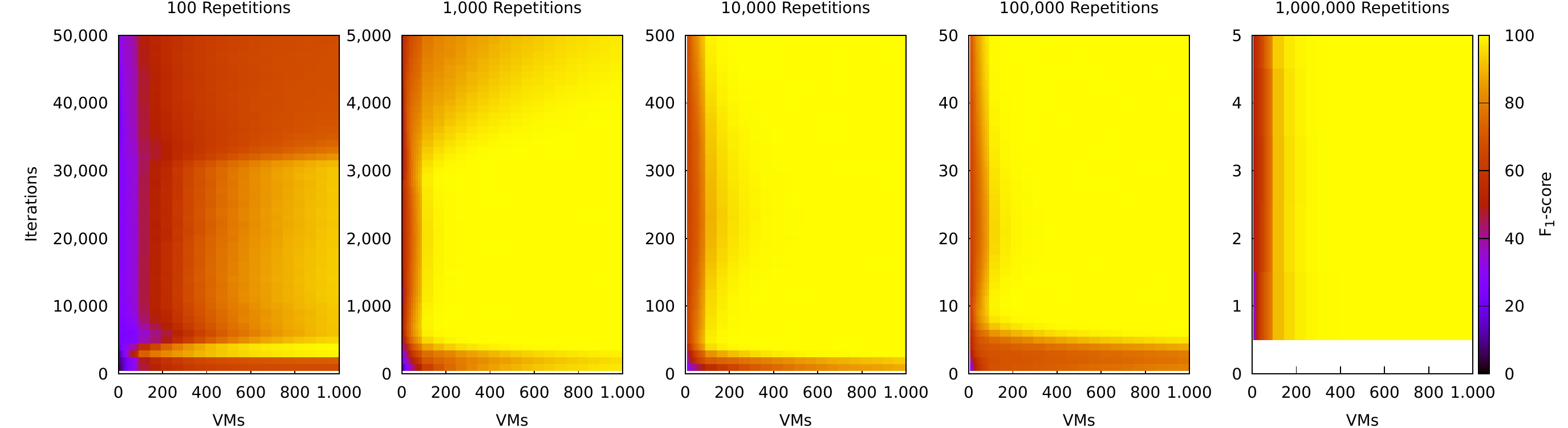}
  \caption{Average $F_1$-score with T-Test}
  \label{fig:heatmapWorkloads}
\end{figure}

\paragraph{Parametrization} 
To identify performance changes reliably, we need to choose the warmup iteration, measurement iteration, repetition and VM count. Since warmup can not be clearly distinguished from warmed up state by statistical methods \cite{barrett2017virtual}, we always use the same count of warmup and measurement iterations. The same product $repetitions*iterations$ implies the same overall workload executions. For every VM, $iteration$ measurement values are taken. Figure~\ref{fig:heatmapWorkloads} shows the average $F_1$-score of all workload types for different iteration, repetition and VM counts using t-test like recommended by \cite{reichelt2018wospc}. Since we only want 1\,\% false positives, we set the significance level of the t-test to 99\,\%.

\begin{figure}[h!]
  \includegraphics[width=12cm]{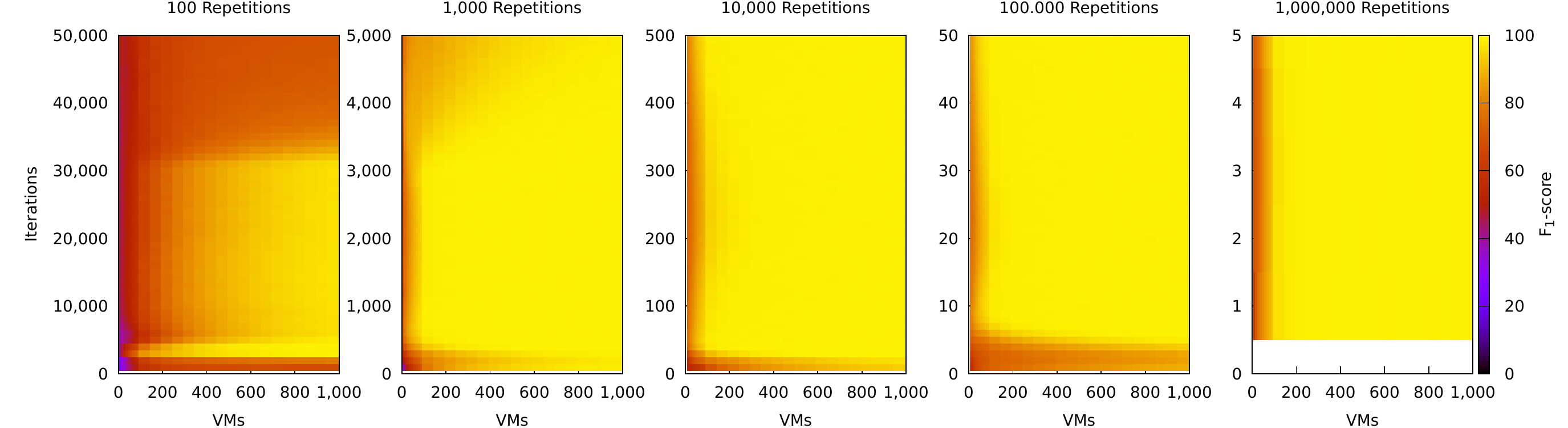}
  \caption{Average $F_1$-score with Confidence Interval Comparison}
  \label{fig:heatmapWorkloadsConfidence}
\end{figure}

\begin{figure}[h!]
  \includegraphics[width=12cm]{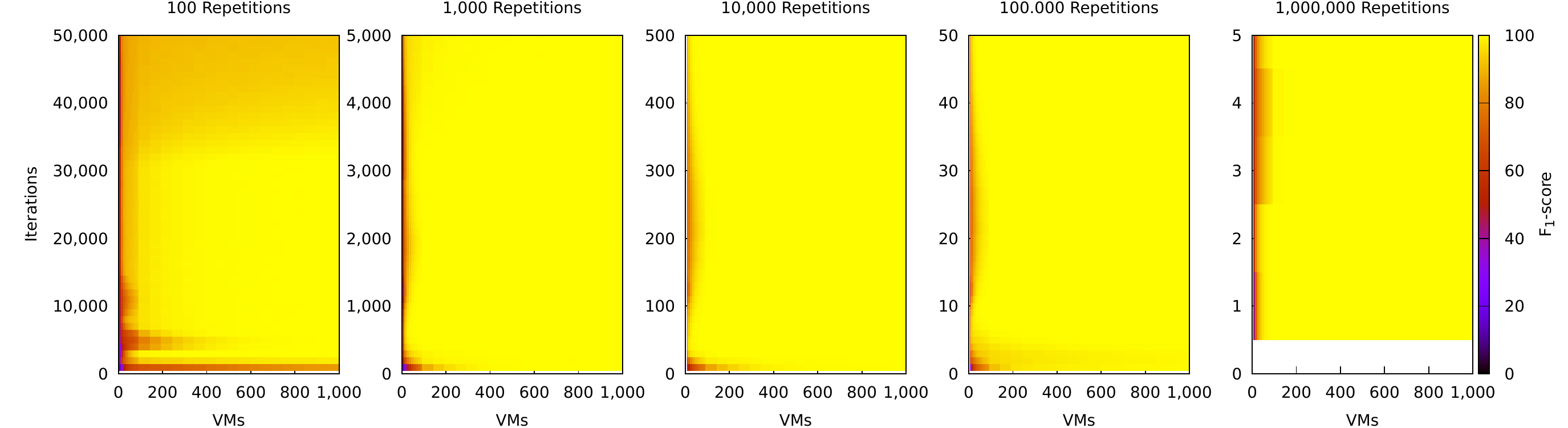}
  \caption{Average $F_1$-score with Mann-Whitney-Test}
  \label{fig:heatmapWorkloadsMannWhitney}
\end{figure}

\begin{figure}[h!]
  \centering
  \includegraphics[width=6cm]{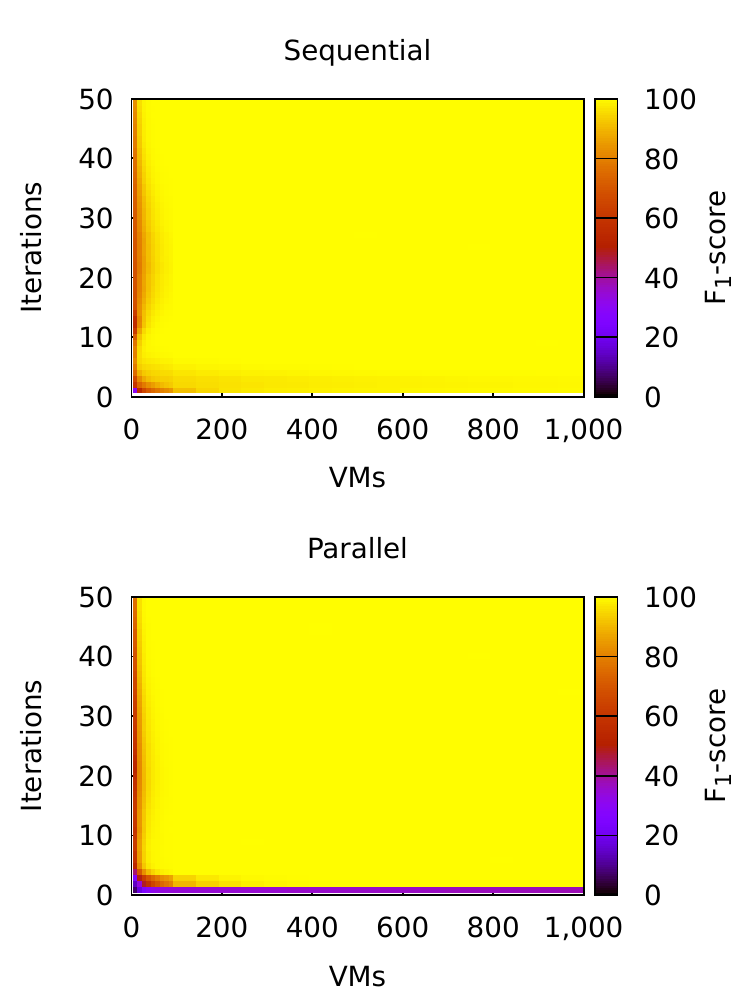}
  \caption{Heatmap of Sequential and Parallel Measurement}
  \label{fig:heatmapParallel}
\end{figure}

\begin{wrapfigure}[31]{l}{0.3\columnwidth}

\end{wrapfigure}

\hspace{-3.5cm}
The $F_1$-score is not only influenced by the measurement configuration, but also by the statistical test. Additionally to t-test, we used confidence interval comparison like recommended by literature \cite{georges2007statistically,kalibera2013rigorous}, and Mann-Whitney-test like recommended by literature \cite{Ding2020, eismann2020microservices}. Figure~\ref{fig:heatmapWorkloadsConfidence} and Figure~\ref{fig:heatmapWorkloadsMannWhitney} show the average $F_1$-score for confidence interval comparison and Mann-Whitney-test. Comparing the heatmaps of the different tests shows that the Mann-Whitney-test is most efficient in identifying correct performance changes in the described setup.
Based on automatic analysis of the $F_1$-scores, 100,000 repetitions, 40\,VMs and 70 iterations using the Mann-Whitney-test is a lower boundary for reliable identification of performance changes of 0.3\,\% for our workload types. We therefore use this as a base for further measurements. For our workloads, the measurement takes 68.7 minutes on average. For bigger workloads, the measurement time might increase.
For measuring workloads with a size of 300, triggering garbage collection increases the execution duration by a factor of 50, but does not significantly decrease the standard deviation. Therefore, we do not execute garbage collection between the iterations.

\paragraph{Parallelization}

Parallelizing performance measurements in\-creas\-es measurement accuracy in cloud environments \cite{bulej2020duet}. Since unit-test-sized performance benchmarks often contain sequential workloads to examine which algorithm or JVM feature usage is faster, parallelization is likely to not reduce accuracy for our performance measurement. We executed the measurement of sequential and parallel execution of the workloads for 100,000 repetitions and performed the Mann-Whitney-test. Figure~\ref{fig:heatmapParallel} shows the results.
It shows that parallelization further increases the $F_1$-score for all workload types. By automated analysis of the $F_1$-scores, we find that 30 VMs and 49 iterations using 100,000 repetitions and the Mann-Whitney-test is a lower boundary for reliable identification of performance changes of 0.3\,\% for our workload types. This measurement takes on average 22.8 minutes.

\begin{figure}[h]
  \centering
  \includegraphics[width=8cm]{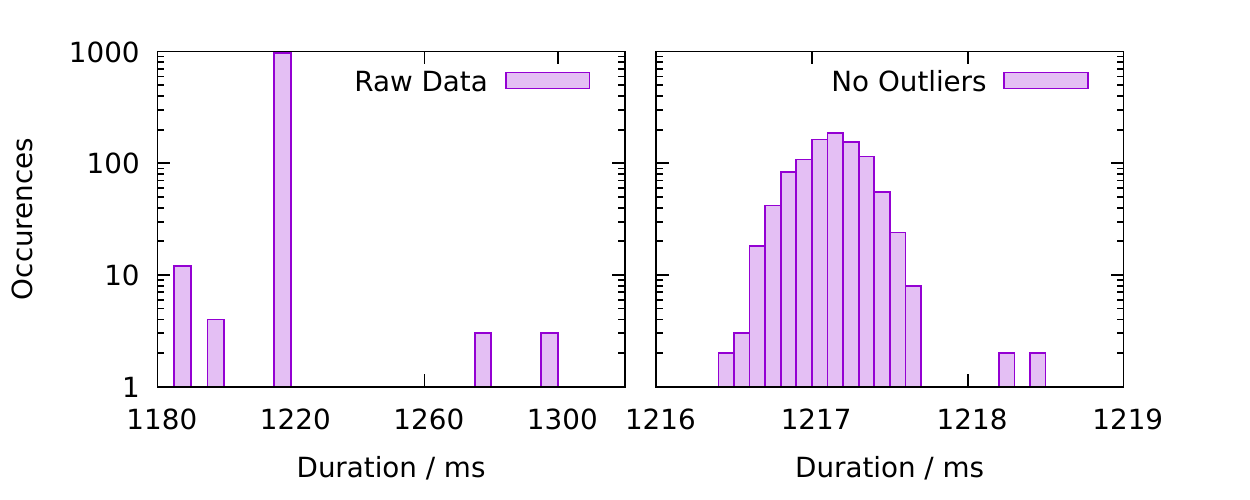}
  \caption{Histogram of Add Durations}
  \label{fig:addHistogram}
\end{figure}

\paragraph{Outlier Removal} 
Figure~\ref{fig:addHistogram} shows the histogram of VM means of \lstinline'Add' workload measurements with 100,000 repetitions. Besides the expected measurement values, which vary around 1,217\,ms, we see some outliers. Since these may change the mean and standard deviation heavily, and these are the main input values for the t-test, we check whether removal of the outliers improves the measurement results. 
We choose to remove outliers by the Z-score. The Z-score of a measurement value $v$ in a distribution $X$ is $Z=\frac{v-\mu_X}{\sigma_X}$, i.e. the Z-score shows how many standard deviations the point is away from the mean value. The outlier removal is done when a measurement value $v$ has a Z-score above 3.29, so 99.9\,\% of values drawn from a Gaussian distribution are not considered outliers. Our analysis shows that removal of the outlier does not increase the $F_1$-score significantly for our measurement configuration. Therefore, we decided to not remove outliers by default.

\paragraph{Generalizability}
Since we chose typical workloads, a widely used JVM and typical execution hardware, we assume that our measurement process configuration will be applicable in many cases. Still, for different workload types, workload combinations, JVM implementations or measurement hardware, our findings may not hold and the process of parametrization might need to be repeated. This can be done by adding workloads to our repository and/or repeating our measurements on different soft- and hardware environment and repeating the analysis process.\footnote{Repository: \url{https://github.com/DaGeRe/precision-experiments} The calls for the scripts of this paper are described in the dataset.}

\section{Evaluation} 
\label{kap:evaluation}


Our approach aims for identification of performance changes. For the evaluation of our approach, we answer \textbf{RQ~II}: Which share of performance changes at code level can be identified by \Peass and by existing performance regression benchmarks? In this section, we describe our approach for answering this question, present our results and discuss threats to validity.

\subsection{Approach}

To compare \Peass to the usage of existing benchmarks, a test set with source code repositories of projects, a list of the commits which caused performance changes and the root causes of these changes would be ideal. Often, projects maintaining benchmarks only execute them occasionally, e.g. per release, and even if measurement results are present, information about regressions are not saved systematically. Therefore, such a test set unfortunately does not exist. Consequently, we use an existing software and inject performance regressions randomly into the source code. 

The software we use for evaluation needs to contain unit tests and performance benchmarks and it needs to be a software where performance at code level is relevant. Additionally, it needs to be technically suitable for our prototype and therefore be written in Java, versioned by git and contain JUnit tests. Since this applies to the widely used application server Jetty,\footnote{\url{https://github.com/eclipse/jetty.project}} we use it for our evaluation. Realistic source code changes that introduce performance regressions could only be created manually, hence we chose to generate artificial regressions by randomly inserting busy waiting into the source code.\footnote{Regression insert and analysis source code is available: \url{https://github.com/DaGeRe/jetty-evaluation}}


\paragraph{Implementation} 
We generated 1,000 Jetty-commits containing individual regressions.\footnote{Example regression: \url{https://github.com/DaGeRe/jetty-experiments/commit/dcc18d5d880553abe8eeb9b03bac18b72e88c8df\#diff-2c090bafe6fd7ec37872df5d0710caa55d2a6593190e6c8fc3df8eda31566d47R191} The regression is in jetty-util/src/main/java/org/eclipse/jetty/util/thread/\\ReservedThreadExecutor.java  in the method \lstinline'tryExecute'}
Each commit is inserted into an individual branch after the latest commit. To create a regression, we randomly selected a method which is called by a randomly selected benchmark. In the selected node, we insert busy waiting of 5 ns, which will under most circumstances be only two calls to \lstinline'System.nanoTime'. The call tree is discovered by the application monitoring framework \Kieker and Javaparser. Afterwards, we execute both--the benchmarks and our configuration of \Peass--in order to identify the introduced performance regressions. 

In our analyzed commit \lstinline'b56edf', the benchmark \lstinline'TrieBenchmark' fails because parts of the trie variant \lstinline'TernaryTrie' are not implemented. Therefore, we skip the fail on error (\lstinline'-foe') flag that is set in the default execution definition and only run the benchmarks that finish properly. The experiments have been done on a cluster with Intel Xeon E5-2620 @ 2,4\,GHz, CentOS 7.9.2009, 128\,GB RAM and Java 11.0.10  (defaulting to 25\% heap of RAM heap size, i.e. 32\,GB). Our raw results are available.\footnote{Dataset of Jetty Performance Measurement Evaluation: \url{https://zenodo.org/record/6321211\#.Yh-NBhu1JaY}}

\subsection{Results}

By the analysis of our results, we can deduce the degree by which the existing benchmarks and our configuration of \Peass are able to identify performance regressions.

\paragraph{Using existing Benchmarks} We executed the existing benchmarks, which are written using the benchmarking framework \textsc{jmh}\footnote{http://openjdk.java.net/projects/code-tools/jmh/} and configured by a Jenkinsfile, for all regressions. The jetty developers themselves seem to run them only occassionally and not on the Eclipse Jetty Jenkins instance\footnote{https://ci.eclipse.org/jetty/}. Only \PercentJMH of the regressions have been detected by the benchmarks, if we compare their measurement results using two-sided t-test with 99\,\% confidence interval in the default benchmark configuration. All benchmarks take approximately 5 hours for execution; therefore, increasing the count of VM executions for each benchmark is costly. It is very likely that increasing the VM count would help to identify more performance regressions.

%

\paragraph{Using our Configuration of \Peass} To examine the same regressions with \Peass, we identified their call trees and chose one unit test that tests the part of the source. Since each regression is called by many unit tests, execution of all tests would not be feasible. Therefore, we select the change where the relation between calls to the changed method and overall method calls is highest, i.e. where we can expect the highest relative performance change. The measurement was executed with the measurement configuration determined in Section~\ref{kap:detection}.

\PercentPeass of the performance changes were detected. \PercentUnitTest could not be identified since the unit tests' standard deviation was too high in comparison to the introduced regression. \PercentNotCovered of all changes have not been covered by unit tests. 
The measurement execution and preprocessing (creation and analysis of the traces) took on average 4.2 hours, which is a high time consumption for analysis of every commit in the CI, but still faster than the execution of all \textsc{jmh} benchmarks.

\paragraph{Comparison}
The difference in efficiency of \Peass and \textsc{jmh} can be explained by the size of the measured workloads: While the transformed unit test workloads contained on average 14,713 method calls (14,628 calls for the correct change detections and 14,902 calls for the incorrect change detections), the benchmarks contained on average 689,150 method calls (15,638 for the correct change detections and 711,415 for the incorrect change detections). This does not mean that a high share of changed method calls on the overall method call count implies easy to measure performance changes: For the transformed unit tests, on average the changed method invocations made up 18.3\,\% of all method invocations (successful change identification: 18.5\,\%, unsuccessful: 18.1\,\%). For the reused \textsc{jmh} benchmarks, 17.5\,\% of all method invocations of the benchmarks are calls of the changed method (successful change identification: 11.0\,\%, unsuccessful: 17.7\,\%). 

Due to the high share of performance regressions identified by \Peass, we conclude that \Peass is able to identify additional performance regressions with low manual effort. Still, performance regressions caused by concurrent usage of software or by different API usage from unit tests will not be detected. Therefore, definitions of benchmarks and load tests will still be necessary. 

\subsection{Threats to Validity}

\paragraph{External Validity} We validated \Peass on artificial performance regressions of the Jetty server. It is unclear whether our results are generalizable for \textit{other examined software} or for \textit{other performance regression} than our artificial regressions, e.g. regressions only occurring with parallel usage. Since there is no defined performance regression set for a software repository, we needed to use artificial regressions. Since we needed to compare our findings to other regressions, we were only able to examine the code which is tested by Jetty's own regression benchmarks. In the future, the generalizability of our results might be increased by examining more software.

Furthermore, it is not possible to research how many \textit{false alarms} about performance regressions are created by \Peass. The definition of the relevance of performance regressions to a software would require expert knowledge on this software and therefore require a separate case study.

\paragraph{Internal Validity} It may be possible that non-deterministic effects influence our performance measurements. Since we applied the regression benchmarks and \Peass to 1,000 regressions, this is very unlikely.

\section{Related Work} 
\label{kap:related}

For the detection of performance changes, there are works focussing on \textit{statistically reliable change detection}, on practical implementations of \textit{regression benchmarking}, on \textit{mining repositories for performance changes} and on \textit{root cause analysis}.

\subsection{Statistically Reliable Change Detection} 
Georges et al. \cite{georges2007statistically} perform a survey of existing work which measures and compares performance values. Thereby, they derive a measurement method which includes the repetition of VM starts and is the base of most recent measurement methods. 
Since measurement is time-consuming, Kalibera et al. \cite{kalibera2013rigorous} provide a method for speeding up the measurement by manually deciding when the warmup is finished. Barrett et al. \cite{barrett2017virtual} define when a steady state is reached by change point analysis. In their measurements, only 43.5\,\% of all benchmarks and VMs reach the steady state. Alghmadi et al. \cite{alghmadi2016automated} propose an approach identifying when measurements are able to stop based on repetitiveness of results, where repetitiveness is seen as difference in terms of the Wilcoxon ranksum statistical test. 
These works focus on performance benchmarks which are written especially for performance measurement, e.g.\ using \textsc{jmh}. In contrary, our work focuses on using usually smaller unit-test-sized workload definitions, which can be obtained by transformation of existing unit tests. 

Kalibera et al. \cite{kalibera2012quantifying} analyze the statistical methods of recent publications. They propose a new method for choosing experiment parameters given a fixed experimentation time and summarizing performance measurements, which includes uncertainty. In contrast to our work, they focus on the statistic analysis, while we focus on configuration of the measurement to reach certain error rates.
Ding et al. \cite{Ding2020} identify performance changes using Mann-Whitney-test, afterwards use Cliff's Delta to get the effect size and then use thresholds for filtering relevant performance changes. In contrast to our work, they focus on selection of unit test cases which are able to measure the performance, instead of a method for unit test execution to identify performance regressions efficiently.
Daly et al. \cite{daly2020use} use, in contrary to the other described works, a list measurement data of several commits to detect performance change points. 
By using change-point analysis, they find the commits that contained performance changes which are persistent over various versions. Thereby, they are able to use less fine-grained measurement data from each individual commit. 
Mühlbauer et al. \cite{muhlbauer2019accurate} also use performance evolution data of several commits, but detect performance changes by modelling the performance evolution using Gaussian processes. 

A recently arising challenge for statistically reliable change detection is the measurement in cloud environments. Laaber et al. \cite{laaber2019software} find that Wilcoxon ranksum test is most efficient, and that at least 20 cloud instances are required for detection of performance changes of 10\,\% mean difference. Bulej et al. \cite{bulej2020duet} find that parallel execution of the same workload in the cloud leads to the best results. He et al. \cite{he2019statistics} state that Kullback-Leibler divergence can be used to determine whether additional measurements are necessary.

\subsection{Regression Benchmarking}
Research has been done to integrate regression benchmarking into build processes. 
There have been studies on open source projects which show that:
\begin{inparaenum}[(1)]
  \item Less than 0.4\,\% of all projects maintain performance benchmarks \cite{stefan2017unit}
  \item If they are maintained, they often have a high variability and are only partially able to identify performance regressions \cite{laaber2018evaluation}.
  \item Developers partially use bad practices in benchmarks, leading to unreliable results \cite{CostoWrongBenchmarkResults}.
\end{inparaenum}
For generic usage, Stochastic Performance Language (SPL) \cite{Bulej2017} defines a method and a tool capable of checking whether a performance change happened based on formulas defining performance requirements. This requires manual workload generation. Rodriguez-Cancio et al. \cite{rodriguez2016automatic} define \textsc{autoJMH}, which automatically generates \textsc{jmh} benchmarks for code snippets. In contrast to our work, they focus on the benchmark generation and not measurement; our measurement configuration could be applied to benchmarks generated by \textsc{autoJMH}. PerfCI \cite{javed2020perfci} is a tool that enables benchmarking of one version and informs the user whether previously specified performance requirements are met. In contrast to our work, they do not compare different versions.

Furthermore, there exist tool-specific inclusions of regression benchmarking into the CI of tools, e.g. \cite{WallerBenchmark} who describe how regression benchmarking was introduced to the \Kieker CI process. Schulz et al. \cite{schulz2021context} discuss how continuous load test generation from production session data can be used to continuously execute load tests of a web application. 

In contrast to our work, these existing works focus on own definitions of performance benchmarks or tests, using their own framework or domain-specific language, while we focus on the measurement process which is suitable for the size of existing unit tests.

\subsection{Mining Repositories for Performance Changes} 
Mining Repositories for performance changes focuses on measurement of code history or extraction of other data to examine performance changes.

\cite{alcocer2015} use Pharo benchmarks to discover performance changes in the history of a repository. Triani et al. \cite{traini2021software} examine performance changes using existing \textsc{jmh}. They use the default configuration of the benchmark, like we did in the evaluation, and compare the measued values by confidence interval comparison and computation of effect sizes like described by Kalibera et al. \cite{kalibera2013rigorous}. They find that certain types of refactoring can increase execution time, e.g. extract class or extract method refactoring. \cite{chen2017exploratory} research changes in RxJava and Hadoop, where they consider changes significant based on a combination of the t-test and the effect size Cohen's $d=\frac{\mu_1-\mu_2}{s}$. 
These studies focus on the identification of changes in the version history; therefore, they presume a performance measurement and analysis method. In contrast, we focus on the measurement method itself and identification of smaller regressions than the aforementioned works.

\subsection{Root Cause Analysis}
Root cause analysis identifies the root cause, i.e.\ the method(s) causing a performance change, by systematic measurements or source code analysis. This might be done using measurement \cite{heger2013,marwedeFailureDiagnosis,alcocer2016}, measurement in combination with workload generation \cite{pradel2014performance,luo2016mining} or source code analysis \cite{chen2020,gu2015change}. By visualization, the understanding of performance changes can be supported \cite{cito2018performancehat,alcocer2019performance}. While measuement methods are partially similiar, these works focus on the identification of the root cause of a performance change, while our work focuses on the identification of the performance change.

\section{Summary} 
\label{kap:summary}

We discussed the statistical boundaries of performance change detection. These are affected by the relation of the performance change size and the measurements standard deviation, i.e. the effect size, and the count of VM starts. We showed that performance changes smaller than the relative standard deviation are only measurable safely with many VMs. This implies that using load tests or measuring the overall performance of big benchmarks is not capable to pinpoint small performance regressions at code level. Small regressions may pile up and decrease the performance of a program. Therefore, methods need to be developed which measure the performance at code level. 

We described the configuration of the method of \Peass, which uses unit-test-sized workloads to identify performance changes. This configuration was obtained by exemplarily measuring artificial workloads and checking whether concrete measurement and analysis configurations are capable of performance change identification. We found that a typical performance change in a unit test, which consists of 300 method calls, may be found using 100,000 repetitions of the workload in 49 iterations, 30 VMs and detection of the change by Mann-Whitney-test.
Based on our results, we described how artificial injected performance regressions in the Jetty application server can be detected. We found that regressions are more easily identifiable using unit tests than using the Jetty benchmark suite.
This work is a step towards \textbf{reliable and efficient automated performance unit test measurement}, which will be able to identify performance changes at code level. Thereby, the widespread approaches of monitoring and load testing or benchmarking can be complemented by another testing stage which requires less additional workload definitions and can be executed earlier in the development cycle.

In future work, there are three challenges to solve:
\begin{inparaenum}[(1)]
  \item Existing unit tests have a high code coverage and therefore tend to test not performance relevant code or test the relevant code with the same workload over and over again. The challenge to \textbf{select the most important test cases for performance measurement} therefore needs to be solved.
  \item Modern systems tend to be highly parallelized. This may render unit test measurements less representative for the overall performance, since parallelized workload has different performance characteristics. The challenge to \textbf{adapt performance measurement at code level to highly parallelized systems} therefore needs to be solved.
  \item While our current work makes it possible to detect the version of the performance change, the concrete method(s) causing the performance change need to be determined manually. The challenge to \textbf{automatically derive the root cause of a performance change} hence needs to be solved. This could be done based on existing root cause analysis techniques \cite{heger2013,marwedeFailureDiagnosis,alcocer2016}. 
\end{inparaenum}

\textbf{Acknowledgments} This work is funded by the German Federal Ministry of Education and Research within the project “Performance Überwachung Effizient Integriert” (PermanEnt, BMBF 01IS20032D).


\bibliographystyle{abbrv}
\bibliography{quellen}

\end{document}